\documentclass[eqsecnum, prd, superscriptaddress, nofootinbib, twocolumn, preprintnumbers]{revtex4-2}
\usepackage{lineno}
\usepackage{amsmath,amssymb}
\usepackage{amsthm}
\usepackage{cases}
\usepackage{physics}
\usepackage{siunitx}
\AtBeginDocument{\RenewCommandCopy\qty\SI}
\usepackage{mathtools}
\usepackage{appendix}
\usepackage{comment}
\usepackage[dvipdfmx]{graphicx}
\usepackage[usenames]{color}
\usepackage{grffile}
\usepackage{mathrsfs}
\usepackage{bm}
\usepackage{url}
\usepackage{ulem}
\usepackage{algorithm}
\usepackage{algorithmic}
\usepackage{natbib}
\usepackage{acronym}
\usepackage{listings}
\usepackage{hyperref}
\usepackage{orcidlink}
\newcommand{\mc}{\mathcal{M}_\mathrm{c}}

\newcommand{\dl}{d_\mathrm{L}}
\newcommand{\un}{\mathrm{n}}
\newcommand{\ur}{\mathrm{r}}

\newcommand{\gw}{\mathrm{gw}}
\newcommand{\bestfit}{\mathrm{bf}}

\newcommand{\etal}{\textit{et al.}}

\begin{document}
\title{Assessing astrophysical foreground subtraction in DECIGO using compact binary populations inferred from the first part of the LIGO-Virgo-KAGRA's fourth observation run}

\preprint{RESCEU-3/26}
\author{Takahiro S. Yamamoto\, \orcidlink{0000-0002-8181-924X}}
\email{yamamoto.s.takahiro@resceu.s.u-tokyo.ac.jp}
\affiliation{Research Center for the Early Universe (RESCEU), Graduate School of Science, The University of Tokyo, Tokyo 113-0033, Japan}
\date{\today}

\begin{abstract}
    Detecting the stochastic gravitational wave background (SGWB) from our Universe under the inflationary era is one of the primary scientific objectives of DECi-hertz Interferometer Gravitational wave Observatory (DECIGO), a space-borne gravitational wave detector sensitive in the $\sim\qty{0.1}{Hz}$ frequency band.
    This frequency band is dominated by the gravitational waves from inspiraling compact object binaries.
    Subtracting these signals is necessary to search for the primordial SGWB.
    In this paper, we assess the feasibility of the subtraction of such binary signals by employing the population model inferred from the latest gravitational wave event catalogue of the LIGO-Virgo-KAGRA Collaboration.
    We find that the projection scheme, which was originally proposed by Cutler~\& Harms (2005), is necessary to reduce the binary signals to the level where DECIGO can detect the primordial background.
\end{abstract}

\maketitle
\acresetall

\acrodef{GW}{gravitational wave}
\acrodef{SNR}{signal-to-noise ratio}
\acrodef{PSD}{power spectral density}
\acrodef{ISCO}{inner-most stable circular orbit}
\acrodef{BBH}{binary black hole}
\acrodef{BNS}{binary neutron star}
\acrodef{BH}{black hole}
\acrodef{NS}{neutron star}
\acrodef{CBC}{compact binary coalescences}
\acrodef{SGWB}{stochastic gravitational wave background}
\acrodef{LVK}{LIGO-Virgo-KAGRA}
\acrodef{GWTC}{Gravitational-Wave Transient Catalog}

\section{Introduction}

It is believed that our Universe experienced the accelerated expansion, called \textit{inflation}, at its very beginning.
Although there are many mechanisms to realize the inflationary expansion, no concrete conclusion has been achieved yet.
One of the keys to unveiling the inflationary universe is the primordial \ac{SGWB} that originates from the quantum fluctuation of the tensor perturbation.
In this work, we consider DECi-hertz Interferometer Gravitational wave Observatory (DECIGO)~\cite{Kawamura:2020pcg}, which is a space-based laser interferometer, as a probe of the primordial \ac{SGWB}.
DECIGO has the sensitivity at the frequency band from \qty{0.01}{Hz} to \qty{10}{Hz}, and their configurations can be optimized for searching primordial \ac{SGWB} (see Sec.~\ref{subsec: signal and detector}).

One of the difficulties for DECIGO to detect the primordial \ac{SGWB} is the presence of a lot of compact binaries, such as \acp{BNS} and \acp{BBH}.
\Acp{GW} they emit overlap each other and form the astrophysical \ac{SGWB} whose power is $O(10^{4-5})$ times larger than the primordial one.
Several strategies have been proposed to address this issue so far.
Biscoveanu~\etal~\cite{Biscoveanu:2020gds} proposes the Bayesian approach to search for the primordial \ac{SGWB} in the presence of the astrophysical foreground.
Their algorithm simultaneously estimates the parameters characterizing the primordial \ac{SGWB} and those characterizing the \ac{BBH} foreground.
The global fit~\cite{Cornish:2005qw,Littenberg:2020bxy} offers another approach in the context of Laser Interferometer Space Antenna (LISA)~\cite{LISA:2024hlh}, the space-borne interferometers aiming for \acp{GW} in millihertz frequency band.
It is expected to observe a lot of signals overlapping each other in the time domain.
The global fit estimates the source parameters for all binaries simultaneously and efficiently, which can be utilized to remove the foreground and search for the primordial \ac{SGWB}~\cite{Rosati:2024lcs}.
Cutler~\& Harms~\cite{Cutler:2005qq} takes the approach to subtract the best-fit waveform of each binary signal.
Many signals from compact binaries can be identified as individual binaries if they have enough amplitudes for DECIGO to detect.
Once we detect them, we can estimate their source parameters and reconstruct their waveform.
By subtracting the reconstructed waveforms from the strain data, we can remove the astrophysical \ac{SGWB}.
The remaining components originate from parameter estimation errors and from binaries not detected as individual events.
Zhu~\etal~\cite{Zhu:2012xw} takes this approach to assess the foreground with the ground-based detectors.

In this paper, we assess how effectively the subtraction scheme could suppress the astrophysical \ac{SGWB}.
We follow Cutler~\& Harms~\cite{Cutler:2005qq} and Rosado~\cite{Rosado:2011kv} for the estimation of the astrophysical \ac{SGWB}.
The population model and the mass spectrum are taken from those inferred from the fourth version of \ac{LVK} Collaboration's event catalog, \ac{GWTC}-4.
In Sec~\ref{sec: method}, we explain the settings for our calculation, such as the signal and detector noise model, the population of \acp{BBH} and \acp{BNS}, and the procedure for calculating the foreground.
As we explain in section~\ref{subsec: sgwb}, the astrophysical foreground can be decomposed into three components.
We calculate each of them to explicitly show how these three components behave, while assessing the subtractable components is enough to discuss the feasibility of the foreground subtraction.
After presenting the results in Sec~\ref{sec: result}, we conclude the paper in Sec~\ref{sec: conclusion}.
Appendix~\ref{appendix} explains the subtraction strategy and its details for estimating the foreground.
The code used to analyze the data is available from~\cite{sourcecode}.

\section{Method}
\label{sec: method}

\subsection{Signal model and detector setup}
\label{subsec: signal and detector}

\begin{figure}[th]
    \centering
    \includegraphics[width=0.9\linewidth]{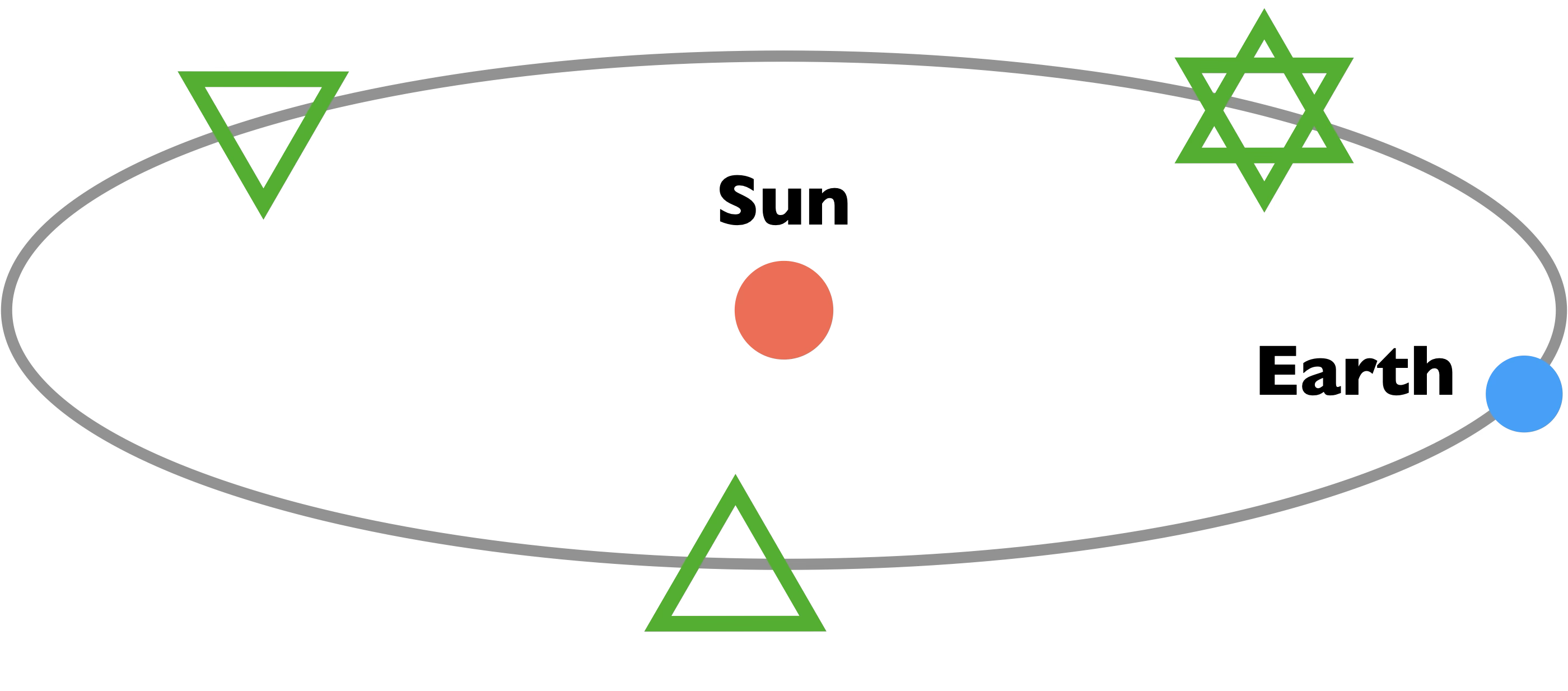}
    \caption{Configuration of DECIGO. In this work, we assume two triangle-shaped consterations and one six-pointed star consteration.}
    \label{fig: decigo configuration}
\end{figure}

DECIGO is a space-based laser interferometer that was originally proposed by Seto \etal~\cite{Seto:2001qf}.
The configuration is shown in Figure.~\ref{fig: decigo configuration}.
Its basic unit consists of three satellites that are located at the corners of an equilateral triangle with the side length of \qty{1000}{km}.
The unit orbits around the Sun with maintaining its triangular formation.
Each satellite has two free-falling mirrors that serve as test masses.
Each pair of two satellites forms a Fabry-Perot cavity and measures the changes in arm lengths induced by gravitational waves.
A triangle-shaped laser interferometer is equivalent to two independent L-shaped interferometers that are located at the same place and have orientations with 45 degree offset.
The ultimate configuration of DECIGO consists of four units.
Two of them are located at different positions on the heliocentric orbit.
The other two units are colocated and have the opposite direction, indicated by the six-pointed star in the Figure~\ref{fig: decigo configuration}.
The star-like configuration is useful for the correlation analysis.
The \ac{SGWB}, whose realization is determined by a random process, would be detected by taking the correlation between two or more independent detectors.
This is a reason why the star-like configuration is proposed.
From the star-like configuration, we can calculate the correlations between two L-shaped interferometers with their arms aligned.
As a result, the star-like configuration provides us with two correlations.
In the following, we assume that we have eight independent L-shaped interferometers and two streams of correlations.
We denote them by
\begin{equation}
    N_\mathrm{det} = 8\,,\quad
    N_\mathrm{corr} = 2\,.
\end{equation}

\begin{figure}[t]
    \centering
    \includegraphics[width=0.8\linewidth]{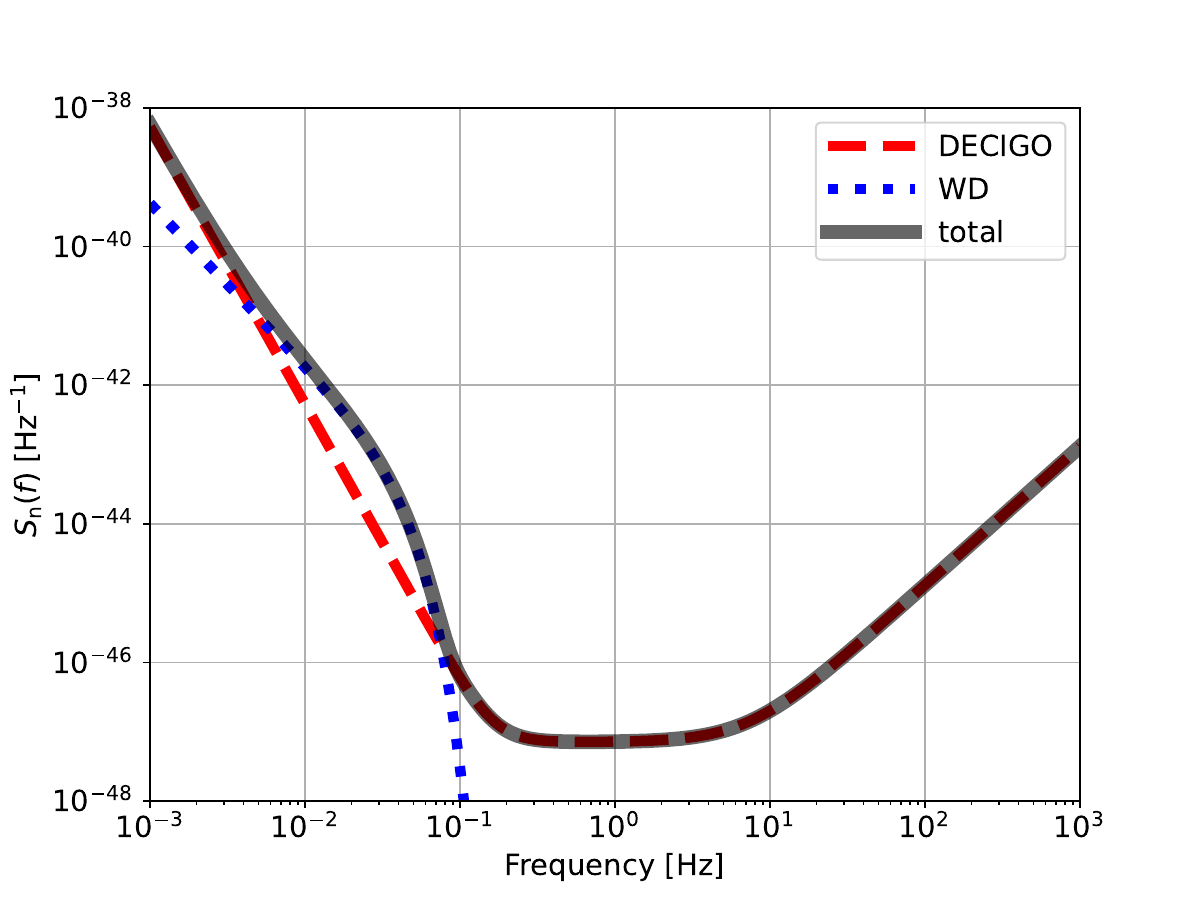}
    \caption{
    Total \ac{PSD} of DECIGO.
    Red dashed line shows the \ac{PSD} of DECIGO's instrumental noise, which is taken from Yagi~\& Seto~\cite{Yagi:2011wg}.
    Blue dotted line represents the foreground originating an ensemble of white dwarf binaries.
    The total \ac{PSD}, the sum of the DECIGO's \ac{PSD} and the white dwarf foreground, is shown by the black solid line.
    }
    \label{fig: decigo psd}
\end{figure}

The noise level of the laser interferometers is characterized by the \ac{PSD}, denoted by $S_\un(f)$.
For the \ac{PSD} of the instrumental noise, we use the formula shown in~\cite{Yagi:2011wg},
\begin{widetext}
\begin{equation}
    S^\mathrm{D}_\un(f) = \num{7.05e-48} \pqty{1 + \pqty{\frac{f}{f_\mathrm{p}}}^2}
    + \num{4.8e-51} \pqty{\frac{f}{\qty{1}{Hz}}}^{-4} \frac{1}{1 + (f/f_\mathrm{p})^2} + \num{5.33e-52} \pqty{\frac{f}{\qty{1}{Hz}}}^{-4}\ \unit{Hz^{-1}}\,,
\end{equation}
\end{widetext}
with $f_\mathrm{p} = \qty{7.36}{Hz}$.
It is known that \acp{GW} from extra-galactic white dwarf binaries would dominate the low frequency band~\cite{Farmer:2003pa, Barack:2003fp}.
We use the fitting formula
\begin{align}
    S^\mathrm{WD}_\un(f) &= \num{4.2e-47} \pqty{\frac{f}{\qty{1}{Hz}}}^{-7/3}\\
    \quad &\times\exp{-2 \pqty{\frac{f}{\qty{5e-2}{Hz}}}^2} \unit{Hz^{-1}}\,,
\end{align}
which is given by~\cite{Barack:2003fp}.
In total, we get the DECIGO's noise spectrum by
\begin{equation}
    S_\un(f) = S^\mathrm{D}_\un(f) + S^\mathrm{WD}_\un(f)\,.
\end{equation}
Figure~\ref{fig: decigo psd} shows the total noise spectrum.
Here, we ignore the galactic component by assuming that all signals from galactic white dwarf binaries can be identified as the individual events and subtracted them from the strain data. This assumption can be justified from the number of white dwarf binaries with the frequency above \qty{1}{mHz}. Under the assumption that a white dwarf binary is a progenitor of a Type Ia supernova, we roughly estimate that there are $\sim \num{e+3}$ white dwarf binaries with the orbital frequency above $\sim \qty{1}{mHz}$. These Galactic binaries can be easily identified and subtracted with their extremely large \acp{SNR}. Therefore, we can safely ignore the contribution from the Galactic white dwarf binaries.

In this work, we only use the inspiral part of the binaries.
We use the angle-averaged \ac{GW} signal.
\begin{equation}
    \tilde{h}(f;\mc, z) = \frac{\sqrt{3}}{2} \mathcal{A}(\mc, z) f^{-7/6} e^{i\Psi(f; \mc)}\,,
    \label{eq: tilde h}
\end{equation}
where $\mathcal{A}(\mc, z)$ is the amplitude, and $\Psi(f; \mc)$ is the phase.
The amplitude is defined by
\begin{equation}
    \mathcal{A}(\mc, z) = \frac{1}{\sqrt{30} \pi^{2/3}} \frac{((1+z)\mc)^{5/6}}{\dl(z)}\,,
    \label{eq: amplitude of chirp signal}
\end{equation}
with the chirp mass $\mc = (m_1 m_2)^{3/5} / (m_1 + m_2)^{1/5}$ and the luminosity distance $\dl(z)$.
The explicit expression of $\Psi(f)$ is not given here since we don't use it in this work.
We use the cosmological parameters inferred by the observational results of Planck18~\cite{Planck:2018vyg} via the library \texttt{astropy}.
The \ac{SNR} of the inspiral signal with $N_\mathrm{det}$ detectors is computed in the frequency domain,
\begin{equation}
    \rho(\mc, z) = \sqrt{N_\mathrm{det}} \cdot \left[ 4\int^\infty_0 \dd{f} \frac{|\tilde{h}(f; \mc, z)|^2}{S_\un(f)} \right]^{1/2}\,.
    \label{eq: def of snr}
\end{equation}
Substituting Eqs.~\eqref{eq: tilde h} and~\eqref{eq: amplitude of chirp signal} to Eq.~\eqref{eq: def of snr}, we get
\begin{equation}
    \rho(\mc, z) = \frac{\sqrt{N_\mathrm{det}}}{\sqrt{10}\pi^{2/3}}\cdot \frac{((1+z) \mc)^{5/6}}{d_\mathrm{L}(z)} \bqty{ \int^\infty_0 \dd{f} \frac{f^{-7/3}}{S_\un(f)} }^{1/2}\,.
\end{equation}

\subsection{Population model}
We follow the reference~\cite{LIGOScientific:2025bgj} for calculating the merger rate density.
The merger rate density of \acp{BBH} is modeled by the double power-law, 
\begin{equation}
    R_\mathrm{BBH}(z; \alpha, \beta, z_\mathrm{p})
    = C \frac{(1+z)^{\alpha}}{1 + \pqty{ \frac{1+z}{1+z_\mathrm{p}} }^{\alpha+\beta}}\,,
    \label{eq: merger rate density of bbh}
\end{equation}
with $\alpha=3.2$, $\beta=5.1$, and $z_\mathrm{p}=2.6$.
The constat $C$ is chosen so that Eq.~\eqref{eq: merger rate density of bbh} matches given local merger rate density at the redshift $z=0$. In~\cite{LIGOScientific:2025pvj}, the local merger rate density of \acp{BBH} is estimated \qtyrange{14}{26}{Gpc^{-3} yr^{-1}}. We use
\begin{equation}
    R_\mathrm{BBH}(z=0) = \qty{20}{Gpc^{-3} yr^{-1}}\,,
\end{equation}
as a fiducial value.

For \acp{BNS}, we assume that the \ac{BNS} formation rate $\psi(z)$ follows the star formation rate and is modeled by the same functional form as Eq.~\eqref{eq: merger rate density of bbh} with $\alpha=2.6$, $\beta=3.6$, and $z_\mathrm{p}=2.2$.
The merger rate density $R_\mathrm{BNS}(z)$ is calculated by convoluting the star formation rate $\psi(z)$ and the delay time distribution, that is,
\begin{equation}
    R_\mathrm{BNS}(z) = A \int^\infty_{t_\mathrm{d}^\mathrm{min}} \dd{t_\mathrm{d}} p(t_\mathrm{d}) \frac{\psi(z_\mathrm{f})}{1 + z_\mathrm{f}} \quad (0\leq z \leq 10)\,,
\end{equation}
and $R(z) = 0$ for $z > 10$.
The time delay, $t_\mathrm{d}$, is the interval between a binary formation and the binary merger.
Its distribution $p(t_\mathrm{d})$ is proportional to $t_\mathrm{d}^{-1}$.
The minimum delay is set at \qty{50}{Myr} for \ac{BNS}.
As with the case of \acp{BBH} merger rate, the constant $A$ is chosen so that $R_\mathrm{BNS}(z=0)$ matches given local merger rate density. Reference~\cite{LIGOScientific:2025pvj} estiamte the local merger rate density of \acp{BNS} as \qtyrange{13}{170}{Gpc^{-3} yr^{-1}}. We choose 
\begin{equation}
    R_\mathrm{BNS}(z=0) = \qty{100}{Gpc^{-3} yr^{-1}}\,,
\end{equation}
as a fiducial value.

For the mass distribution, we used the population model inferred from GWTC-4~\cite{LIGOScientific:2025slb, LIGOScientific:2025pvj} .
We employ the \textsc{SimpleUniform} for \ac{BNS}, in which the component masses follow the uniform distribution on $[1.0, 2.5]M_\odot$.
For \ac{BBH}, we used the \textsc{BrokenPowerLaw+TwoPeaks} model.
We obtain the posterior samples from Zenodo repository~\cite{ligo_scientific_collaboration_2025_16911563} and use the median values for the hyperparameters characterizing the mass spectrum.
In this work, we assume that the mass distribution does not depend on the redshift.

\subsection{SGWB}
\label{subsec: sgwb}

The strength of \ac{SGWB} is characterized by the (dimensionless) density parameter, $\Omega_\gw$, which is defined by the energy density of the \ac{SGWB} divided by the critical density.
For the primordial \ac{SGWB}, the simplest inflation model predicts the flat spectrum in DECIGO's frequency band.
In this work, referring to the Planck18's result~\cite{Planck:2018vyg}, we use 
\begin{equation}
    \Omega_\gw(f) \equiv 10^{-16}\,,
    \label{eq: primordial sgwb}
\end{equation}
as a fiducial value.
We calculate the astrophysical \ac{SGWB} generated by the ensemble of compact binaries by following the method used in Rosado~\cite{Rosado:2011kv} and Cutler~\& Harms~\cite{Cutler:2005qq}.
The energy spectrum of \ac{GW} from a binary is given by
\begin{equation}
    \dv{E_\gw}{f_\ur} = \frac{\pi^{2/3}}{3} \frac{G^{2/3} \mc^{5/3}}{f_\ur^{1/3}}\,,
    \label{eq: dE/df}
\end{equation}
where $f_\ur$ is the frequency in the source's rest frame that relates to the frequency in the detector frame by $f_\ur = (1 + z) f$.
The energy spectrum of a binary in logarithmic scale is denoted by
\begin{equation}
    P(f,z) = \bqty{ f_\ur \dv{E_\gw}{f_\ur} }_{f_\ur = (1 + z) f}\,.
    \label{eq: energy power spectrum}
\end{equation}

The astrophysical foreground is calculated by integrating the energy spectrum of a binary over the relevant redshift range.
We define the minimum and the maximum redshift, denoted by $z_\mathrm{min}$ and $z_\mathrm{max}$ respectively, that no binaries are assumed to exist outside of the redshift range from $z_\mathrm{min}$ and $z_\mathrm{max}$.
In this work, we fix them at
\begin{equation}
    z_\mathrm{min} = 0\,,\quad
    z_\mathrm{max} = 10\,.
    \label{eq: zmin, zmax}
\end{equation}
The integration range of the redshift should be determined with accounting for the merger frequencies and the \ac{GW} frequencies shifting as the source distance.
In this work, we use the frequency at the last stable orbit, which is defined by
\begin{equation}
    f_\mathrm{max} = \frac{1}{6\sqrt{6}\pi} \frac{c^3}{G(m_1 + m_2)}\,,
    \label{eq: fmax}
\end{equation}
as a proxy for the merger frequency.
The minimum frequency is the frequency at $T_\mathrm{obs} = \qty{3}{yr}$ before the merger.
It can be written as
\begin{equation}
    f_\mathrm{min} = \bqty{ \frac{T_\mathrm{obs}}{\delta_2} + f_\mathrm{max}^{-8/3} }^{-3/8}
    \simeq \bqty{ \frac{T_\mathrm{obs}}{\delta_2} }^{-3/8}\,,
    \label{eq: fmin}
\end{equation}
with
\begin{equation}
    \delta_2 = \frac{5c^5}{256 \pi^{8/3} (G \mc)^{5/3}}\,.
    \label{eq: delta 2}
\end{equation}
We define the lower limit $z_\mathrm{low}$ and the upper limit $z_\mathrm{upp}$ by
\begin{equation}
    z_\mathrm{upp}(f) = 
    \begin{dcases}
    z_\mathrm{max} & f \leq \frac{f_\mathrm{max}}{1 + z_\mathrm{max}}\,,\\
        \frac{f_\mathrm{max}}{f} - 1 & \frac{f_\mathrm{max}}{1 + z_\mathrm{max}} < f < \frac{f_\mathrm{max}}{1 + z_\mathrm{min}} \,,\\
        z_\mathrm{min} & \frac{f_\mathrm{max}}{1 + z_\mathrm{min}} \leq f\,,
    \end{dcases}
    \label{eq: zupper}
\end{equation}
and
\begin{equation}
    z_\mathrm{low}(f) = \begin{dcases}
        z_\mathrm{max} & f \leq \frac{f_\mathrm{min}}{1 + z_\mathrm{max}}\,,\\
        \frac{f_\mathrm{min}}{f} - 1 & \frac{f_\mathrm{min}}{1 + z_\mathrm{max}} < f < \frac{f_\mathrm{min}}{1 + z_\mathrm{min}} \,,\\
        z_\mathrm{min} & \frac{f_\mathrm{min}}{1 + z_\mathrm{min}} \leq f\,.
    \end{dcases}
    \label{eq: zlower}
\end{equation}
They depend on the source parameters, particularly on the chirp mass.
We do not indicate the dependency on the source parameter to make the notation simple.
The density parameter $\Omega_\mathrm{fg}(f)$ of the astrophysical foreground is calculated by 
\begin{equation}
    \Omega_\mathrm{fg}(f) = \frac{1}{\rho_\mathrm{c} c^2}
    \left\langle \int^{z_\mathrm{upp}}_{z_\mathrm{low}} \dd{z} \frac{R(z)P(f,z)}{H(z) (1 +z)^2} \right\rangle_\mathrm{s}\,,
\end{equation}
where $\left\langle \cdot \right\rangle_\mathrm{s}$ implies the average over the source parameters, $H(z)$ and $\rho_\mathrm{c}$ are the Hubble parameter and the critical density, respectively.

\subsection{Foreground subtraction and its residual}
\label{subsec: subtraction and residual}

\begin{figure}[t]
    \centering
    \includegraphics[width=1.0\linewidth]{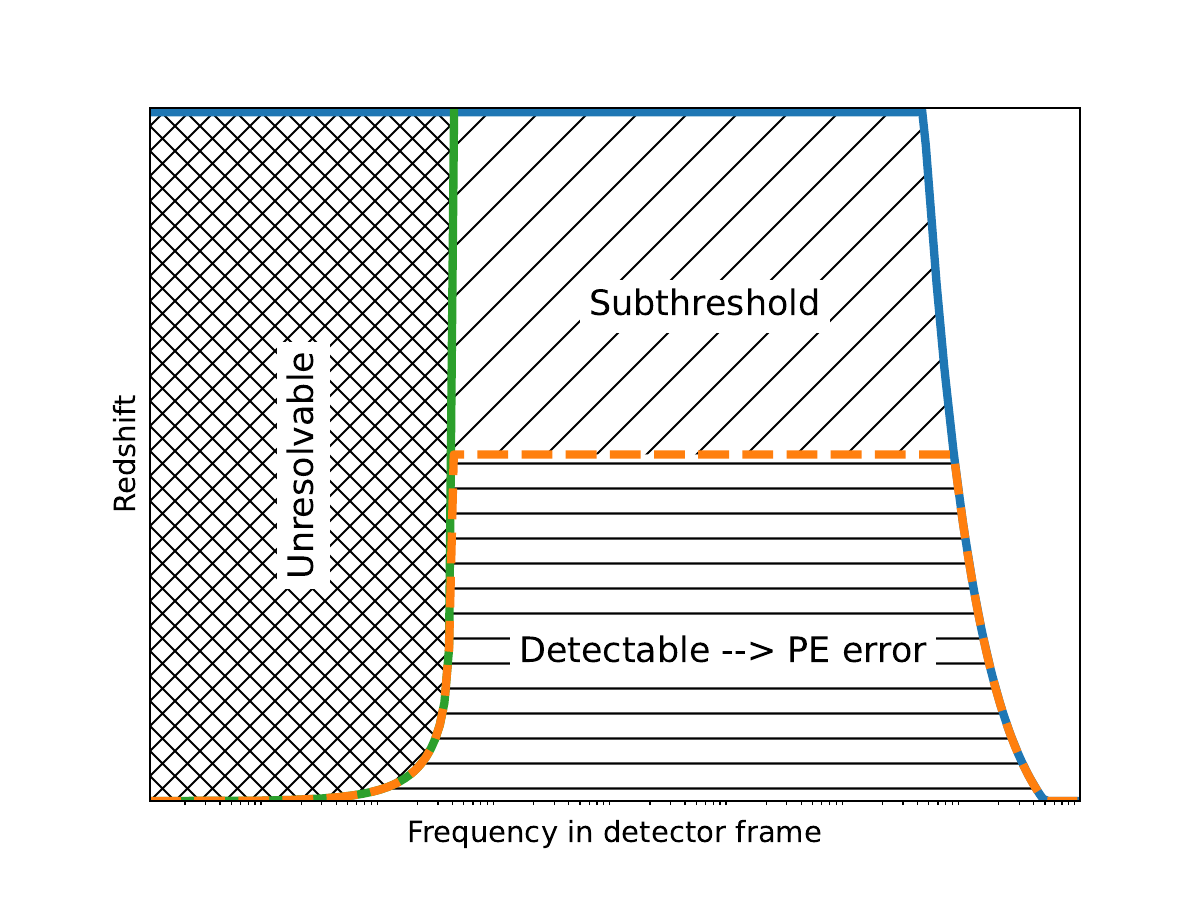}
    \caption{\label{fig: zf_diagram}
    Three components of the foreground.
    The cross-hatched region is the unresolvable part contributing to $\Omega_\mathrm{unres}$.
    Binaries in the region hatched by the diagonal lines generate the subthreshold component, $\Omega_\mathrm{subth}$.
    The region hatched by the horizontal lines corresponds to the detectable part, in which binaries can be detected as individual events.
    The errors in parameter estimation for these binaries contribute $\Omega_\mathrm{err}$.
    The blue line shows $z_\mathrm{upp}$ defined by Eq~\eqref{eq: zupper}.
    The green line is $\mathfrak{z}$ that is defined by Eq.~\eqref{eq: mathfrak z}.
    The orange dashed line is the upper bound of the detectable region.
    The right blank region indicates that there are no \acp{BNS} because it is a frequency band above the last stable orbit, and we ignore the merger and the post-merger part.}
\end{figure}

In this subsection, we explain the strategy for subtracting the astrophysical foreground and our estimate of the residuals that will remain in the strain data after subtraction.
These residuals contribute a noise source in the sense that we cannot detect the primordial \ac{SGWB} if the residual dominates.

We assume two strategies for subtracting the astrophysical foreground.
In the first strategy, we search the signals from binaries, estimate their source parameters, and subtract the waveforms with the best-fit parameters from the strain data.
The second strategy employs the projection scheme originally proposed by Cutler~\&~Harms~\cite{Cutler:2005qq}.
After subtracting the best-fit waveforms, as we did in the first strategy, the residuals will remain in the strain data. These residual has a component that is tangential to the waveform hypersurface in the high-dimensional signal space. By using the projection operator, defined by Eq.~\eqref{eq: projection op}, the tangential components can be suppressed, leading to further suppression of the subtraction errors.

The residual can be decomposed into three parts: the unresolvable part, the subthreshold part, and the parameter estimation (PE) error part.
Figure~\ref{fig: zf_diagram} shows that the redshift and frequency plane is divided into the subregions corresponding to the above three components.
The unresolvable part mostly appears in the low-frequency band.
In general, the lower the orbital frequency of the binary, the slower its frequency evolution is.
The binary with the lower orbital frequency tends to stay in the same frequency bin for a long time.
If two or more binaries coexist within one frequency bin during the entire observational period, you cannot resolve and separate them.
This is the origin of the unresolvable part.
The subthreshold part consists of binaries with \ac{SNR} below the detection threshold.
Binaries with \ac{SNR} above the threshold can be detected as individual events and will be subtracted.
The residuals from estimation errors accumulate to form the PE error component.
In the following, we evaluate each residual component to clarify which components may potentially pose difficulties in the detection of the primordial \ac{SGWB}.

First, we estimate the unresolvable part denoted by $\Omega_\mathrm{unres}$.
To estimate the unresolvable part, we need the information about how many binaries are within a given frequency resolution at a given redshift.
The number of binaries within the redshift $z$ and located in the frequency range from $f$ to $f+\Delta f$ is
\begin{equation}
    \mathcal{N}(f, \Delta f, z) = \left\langle \int^z_{z_\mathrm{low}} \dd{z'} \tau_\mathrm{e}(f, \Delta f, z'; \theta) \dot{N}(z'; \theta) \right\rangle_\mathrm{s}\,.
    \label{eq: num of binaries within a freq bin}
\end{equation}
$\tau_\mathrm{e}(f, \Delta f, z;\theta)$ is the time interval where a binary with the source parameter $\theta$ at the redshift of $z$ spends from the frequency of $f$ to $f + \Delta f$.
It is calculated by
\begin{equation}
    \tau_\mathrm{e}(f, \Delta f, z; \theta)
    = \delta_2 (1 + z)^{-8/3} \bqty{f^{-8/3} - (f + \Delta f)^{-8/3}}\,,
\end{equation}
with $\delta_2$ defined by Eq.~\eqref{eq: delta 2}.
$\dot{N}(z; \theta)$ is the number of binaries with the source parameter $\theta$ at the redshift $z$.
At a given frequency $f$ and a frequency resolution $\Delta f$, we can calculate the redshift $\mathfrak{z}$ below where one binary is expected to exist.
By this definition, the redshift $\mathfrak{z}$ is calculated by solving the equation
\begin{equation}
    \mathcal{N}(f, \Delta f, \mathfrak{z}) = 1\,.
    \label{eq: mathfrak z}
\end{equation}
The unresolvable part can be calculated by
\begin{equation}
    \Omega_\mathrm{unres}(f) = \frac{1}{\rho_\mathrm{c}c^2} 
    \left\langle \int^{z_\mathrm{upp}}_{z_1} \dd{z} \frac{R(z)P(f,z)}{(1 + z)^2 H(z)} \right\rangle_\mathrm{s}\,,
\end{equation}
with $z_1 \coloneqq \min [\mathfrak{z}, z_\mathrm{low}]$.
Accounting for the detector motion inducing the Doppler modulation in signal frequencies, we may still resolve the binaries even when they are located within a single frequency bin. In this work, we naively set the criterion that binaries within one frequency bin can not be resolved.

Second, we estimate the PE error part.
A binary detectable as an individual event satisfies the condition of
\begin{equation}
    \rho(\mc, z) \geq \rho_\mathrm{th}\,,
\end{equation}
with the detection threshold $\rho_\mathrm{th}$.
The redshift $z_\mathrm{th}$ is naturally defined by $\rho(\mc, z_\mathrm{th}) = \rho_\mathrm{th}$.
In this work, we use the following considerations to assess the contribution from the parameter estimation errors.
The assumption that the waveform does not have systematics due to mismodeling is implicitly made.
We also do not account the calibration erros that can lead the systematic bias in the parameter estimation.
We only account statistical errors in the parameter estimation.
We approximate the contribution to $\Omega_\mathrm{err}$ from each binary by
\begin{equation}
    \dv{E_\gw}{\ln f_\ur} \to \dv{E_\gw}{\ln f_\ur} \times \frac{N_\mathrm{p}}{\rho^2}
\end{equation}
where $N_\mathrm{p}$ is the number of source parameters characterizing each binary.
In this work, we set $N_\mathrm{p} = 11$ (component masses, two spin parameters, and seven extrinsic parameters).
This estimation is described in Appendix~\ref{appendix a1: best-fit}.
We estimate $\Omega_\mathrm{err}$ by
\begin{equation}
    \Omega_\mathrm{err}(f) = \frac{1}{\rho_\mathrm{c} c^2}
    \left\langle \int^{z_2}_{z_\mathrm{low}} \dd{z}
    \frac{R(z)P(f,z)}{H(z) (1 + z)^2} \cdot \frac{N_\mathrm{p}}{\rho^2 } \right\rangle_\mathrm{s}\,,
    \label{eq: omega_err}
\end{equation}
with $z_2 \coloneqq \min [\mathfrak{z}, z_\mathrm{th}, z_\mathrm{upp}]$, which is indicated by the orange dashed line in Fig.~\ref{fig: zf_diagram}.

Cutler~\& Harms proposed the scheme suppressing $\Omega_\mathrm{err}$ by projecting out the first order of the parameter deviation from the residual.
The scheme is outlined in the Appendix~\ref{appendix a2: projection}.
After the projection, the residual errors are dominated by the second order of the parameter estimation errors.
Its power is evaluated by
\begin{equation}
    \dv{E_\gw}{\ln f_\ur} \times \frac{|\delta^\mathrm{P} h|^2}{|h|^2}
    \sim \dv{E_\gw}{\ln f_\ur} \times \pqty{\frac{N_\mathrm{p}}{\rho^2}}^2\,.
\end{equation}
Therefore, we estimate the foreground components generated from the subtraction residuals after the projection by
\begin{equation}
    \Omega^\mathrm{P}_\mathrm{err}(f) = \frac{1}{\rho_\mathrm{c} c^2}
    \left\langle \int^{z_2}_{z_\mathrm{low}} \dd{z} \frac{R(z) P(f, z)}{H(z) (1 + z)^2}
    \cdot \pqty{\frac{N_\mathrm{p}}{\rho^2}}^2 \right\rangle_\mathrm{s}\,.
    \label{eq: omega_err_p}
\end{equation}

In the end, we estimate the subthreshold part.
The subthreshold component $\Omega_\mathrm{subth}$ consists of the binaries with the \ac{SNR} less than the detection threshold.
Using the definition of $z_\mathrm{th}$, we estimate the subthreshold component by
\begin{equation}
    \Omega_\mathrm{subth}(f) = \frac{1}{\rho_\mathrm{c} c^2}
    \left\langle \int^{z_3}_{z_2} \dd{z}
    \frac{R(z)P(f, z)}{H(z) (1 + z)^2} \right\rangle_\mathrm{s}\,,
\end{equation}
with $z_3 \coloneqq \min[\mathfrak{z}, z_\mathrm{upp}]$.

We calculate the average over the source parameters by using Monte-Carlo sampling with the sample size of \num{1000}.

\section{Result}
\label{sec: result}

\subsection{Main result}

\begin{figure}[t]
    \centering
    \includegraphics[width=0.9\linewidth]{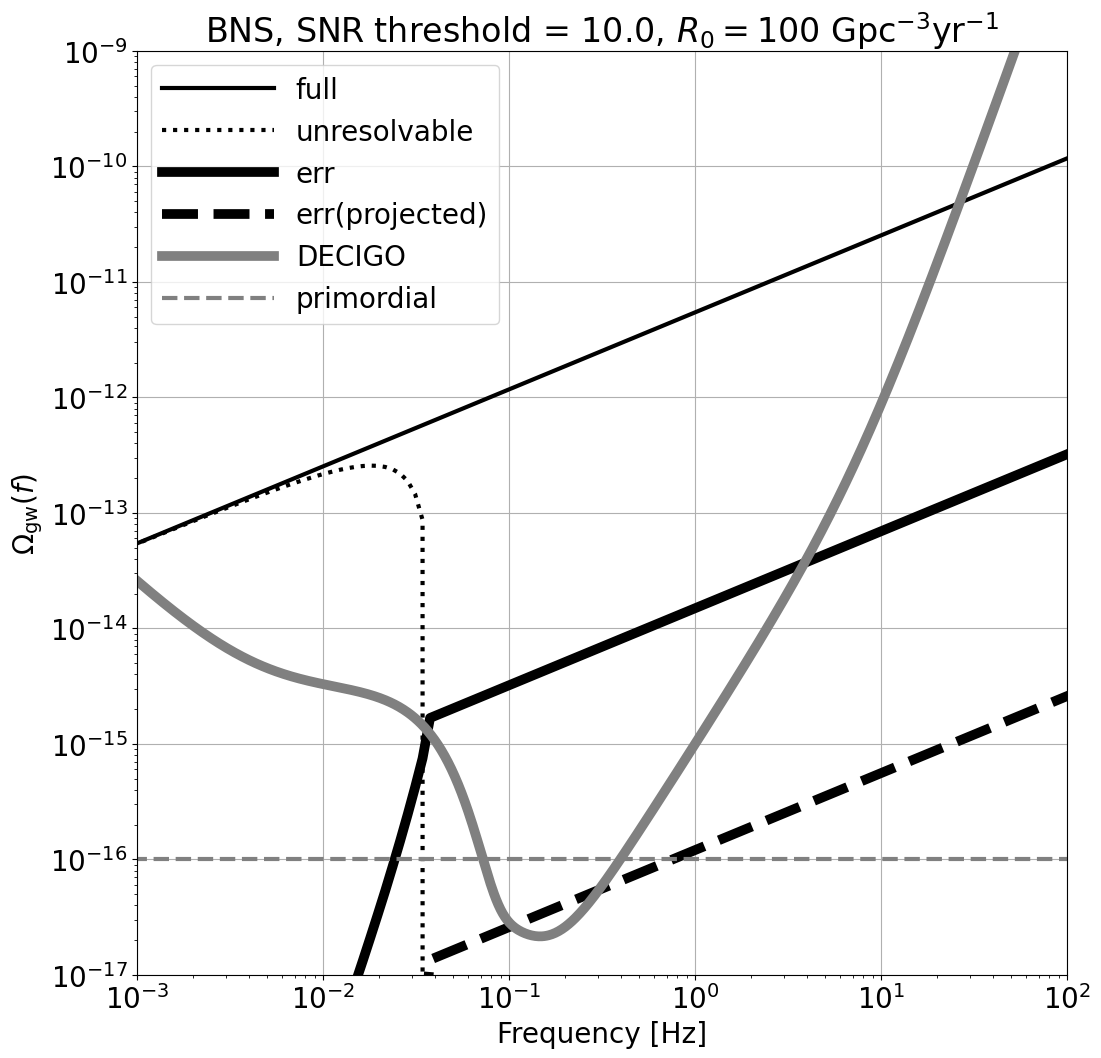}
    \caption{Astrophysical foreground of \acp{BNS} and the subtracted residuals.
    We assume the SNR threshold of 10. The black thin line shows the total foreground.
    The thin dotted line is the unresolvable component. The black thick solid and dashed lines are the subtraction residual and the projected one, respectively.
    The gray solid line is the minimum amplitude of $\Omega_\gw(f)$ that can be detectable by DECIGO with the SNR larger than unity.
    The gray dashed line shows the fiducial value \num{e-16} of the primordial GW background.
    The subthreshold component is too small to be appeared in this plot.}
    \label{fig: bns snrth10}
\end{figure}

Figure~\ref{fig: bns snrth10} shows the estimations of the astrophysical foreground and the subtraction results for the case of \acp{BNS} with $R(z=0) = \qty{100}{Gpc^{-3} yr^{-1}}$ and $\rho_\mathrm{th}=10$.
DECIGO's sensitivity for \ac{SGWB} is also plotted by the gray solid line.
The \ac{SNR} threshold is small enough so that the subthreshold components are well below the primordial components and other foreground components.
In the frequency band below \qty{0.04}{Hz}, the unresolvable components dominate.
Above that region, the subtraction error becomes the dominant one in the foreground.
Without the projection, the foreground by the subtraction errors $\Omega_\mathrm{err}$ is larger than the primordial components by from 1 to 2 orders of magnitude.
This means we cannot detect the primordial components only by subtracting the best-fit waveforms.
The projection scheme reduces it by about two orders of magnitude, making the subtraction errors below the primordial \ac{SGWB}.
This shows that the projection scheme will play a crucial role in detecting the primordial \ac{SGWB} by DECIGO.

\begin{figure}[t]
    \centering
    \includegraphics[width=0.9\linewidth]{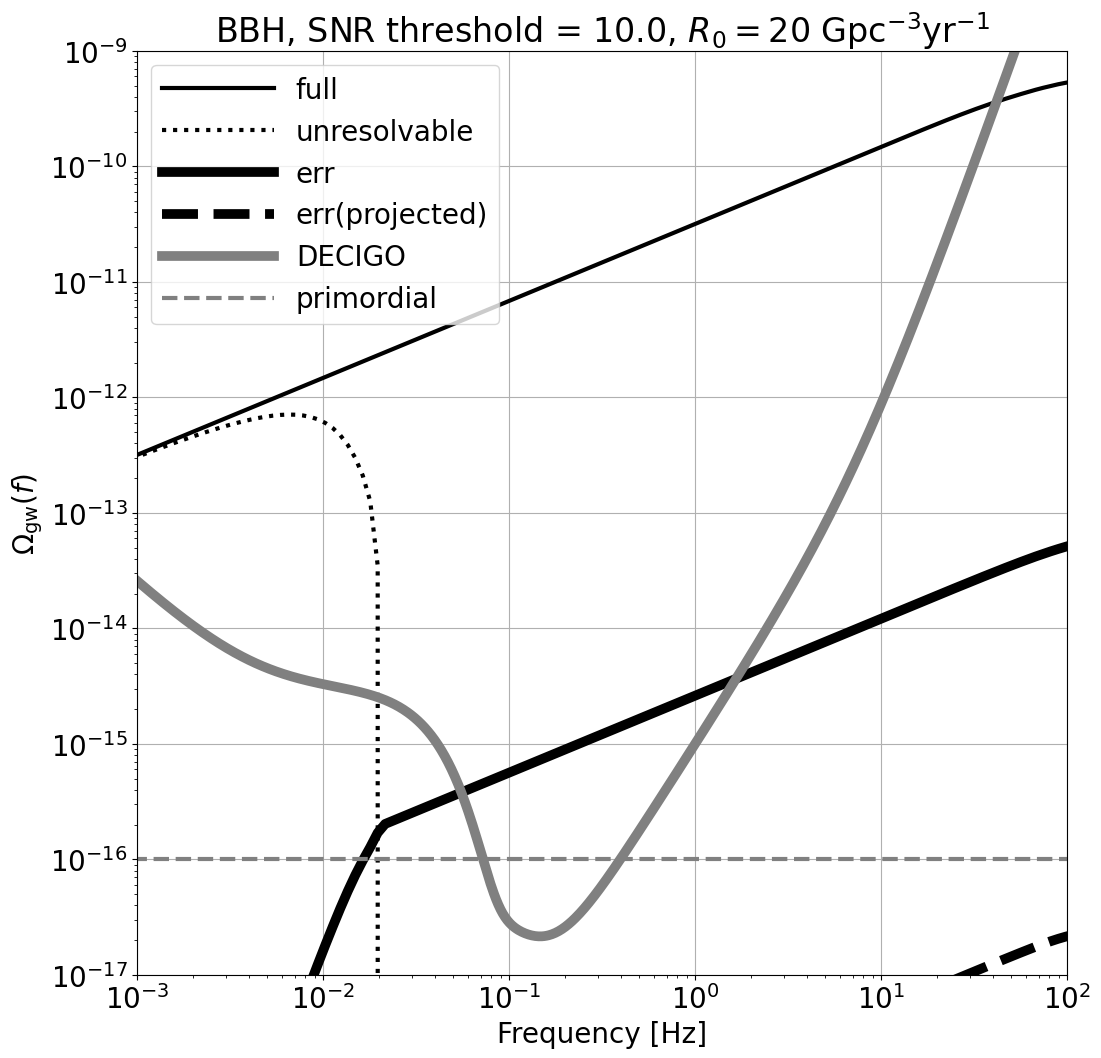}
    \caption{Foreground estimation for \ac{BBH}. The legends are the same as Fig.~\ref{fig: bns snrth10}.}
    \label{fig: bbh snrth10}
\end{figure}

Figure~\ref{fig: bbh snrth10} shows the foreground assessments for \acp{BBH}.
We set $R(z=0) = \qty{20}{Gpc^{-3} yr^{-1}}$ and $\rho_\mathrm{th}=10$.
The amplitude of the total foreground is one order of magnitude larger than that of \acp{BNS}.
Each \ac{BBH} signal of which the foreground consists tends to have a larger amplitude than a \ac{BNS}.
Therefore, DECIGO can detect more distant \acp{BBH} than \acp{BNS}; the subthreshold component will be reduced to a level that it does not affect the detection of the primordial \ac{SGWB}.
The subtraction error in the \ac{BBH} case is also smaller than that of the \ac{BNS} case, though it is not small enough to detect \ac{SGWB}.
We conclude that the projection scheme is also crucial for removing the foreground of \acp{BBH}.

\begin{figure}[t]
    \centering
    \includegraphics[width=0.95\linewidth]{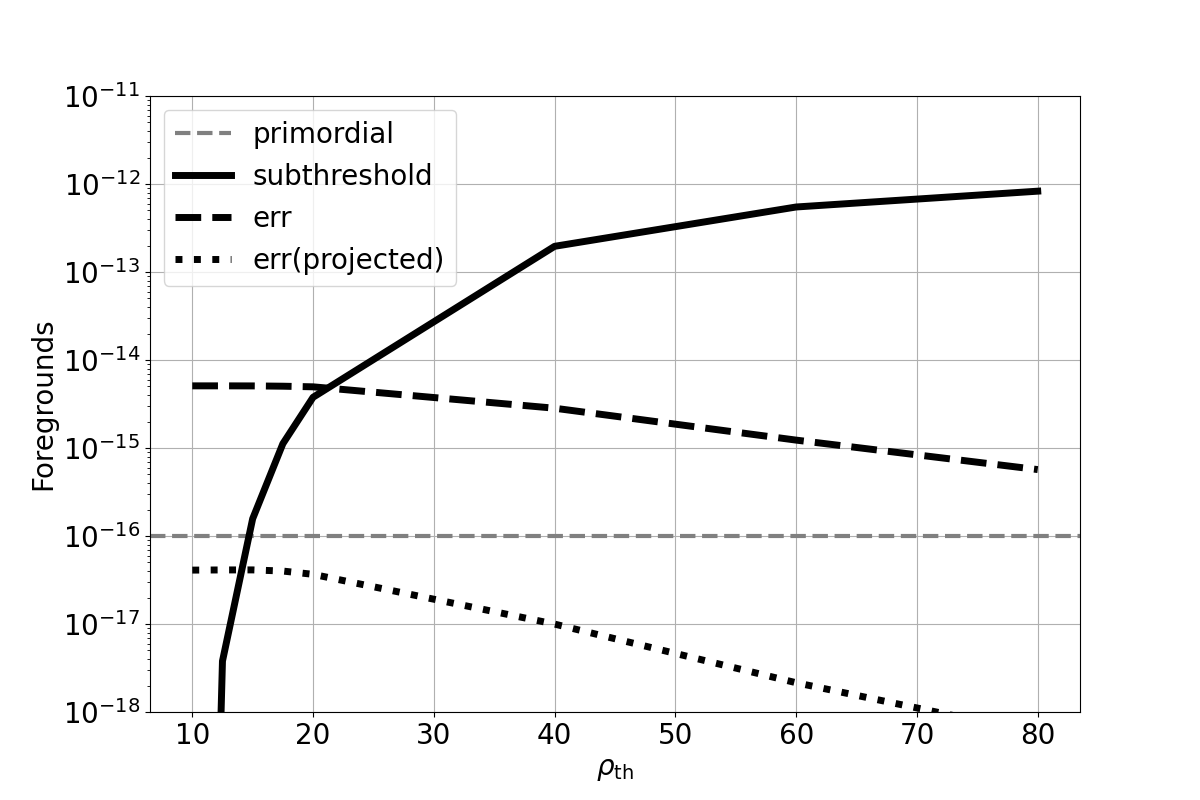}
    \caption{Components of the \ac{BNS} foreground at the frequency of \qty{0.2}{Hz} where DECIGO is the most sensitive to \ac{SGWB}.
    The black dashed and dotted lines are the error components without and with the projection scheme, respectively.
    The black solid line is the subthreshold component.
    The gray horizontal dashed line is the primordial \ac{SGWB}.
    For $\rho_\mathrm{th}=10.0$, the subthreshold component returns \num{0.0} that means that there are no subthreshold events.}
    \label{fig: foregrounds}
\end{figure}

Figure~\ref{fig: foregrounds} shows the components of the foregrounds formed by \ac{BNS} for various \ac{SNR} thresholds.
We fix the frequency at \qty{0.2}{Hz} where DECIGO is the most sensitive to \ac{SGWB}.
As the threshold increases, the subthreshold components become larger while the subtraction error components with and without the projection slowly decrease.
This figure indicates that, in order to subtract the signals from \acp{BNS} and lower the foreground level enough small to detect the primordial \ac{SGWB}~\eqref{eq: primordial sgwb}, we need to set the \ac{SNR} threshold lower than $\sim\num{15}$ and implement the projection scheme.
For \acp{BBH}, we found that all events can be detected even with the threshold of \num{20}, with no subthreshold components.
However, we still need the projection scheme, as the subtraction errors without it are higher than the primordial \ac{SGWB}.

\subsection{Uncertainty in the population model}

We study the impact of the uncertainty in the population model. First, we study the uncertainty in the mass distribution model. Reference~\cite{LIGOScientific:2025bgj} shows various mass spectrum models for both \acp{BBH} and \acp{BNS}. Because only two \acp{BNS} events are discovered by \ac{LVK} Collaboration, the uncertainty in the mass spectrum is relatively large compared to \ac{BBH}. GWTC-4 population paper shows three mass spectrum models: the \textsc{Simple Uniform} model, the \textsc{Peak} model, and the \textsc{Power} model. We perform the same analysis for the \textsc{Peak} model and the \textsc{Power} model as we did for the \textsc{Simple Uniform} model. We find that the change in the mass spectrum models do not significantly affect the foreground levels.
This less susceptibility on the mass spectrum can be understood as follows. As we showed in Sec~\ref{sec: method}, we estimate the error components by Eqs.~\eqref{eq: omega_err} and~\eqref{eq: omega_err_p} where the power of the residuals are assessed only by the number of waveform parameteres and the \ac{SNR}, as shown in Eqs.~\eqref{eq: dh^2/h^2} and~\eqref{eq: dhp^2/h^2}.
The energy power spectrum $P(f, z)$ and the signal \ac{SNR} depend on the chirp mass as $P(f, z) \propto \mc^{5/3}$ and $\rho \propto \mc^{5/6}$ (see Eqs.~\eqref{eq: def of snr}, ~\eqref{eq: dE/df}, and~\eqref{eq: energy power spectrum}). The combination $P(f,z) / \rho^2$ does not depend on the chirp mass. In the calculation of the foreground components, the uncertainty in the mass spectrum affects only on the integration range of the redshift. This is the reason why the uncertainty in the mass spectrum has negligible impact on the foreground estimation.
For \acp{BBH}, the difference in the mass spectrum is not so large compared to the case of \acp{BNS} because \ac{LVK} Collaboration has discovered more than \num{200} \ac{BBH} events in O4a. Therefore, we don't expect that the uncertainty in the \ac{BBH} mass spectrum that is obtained by \ac{LVK} Collaboration significantly changes the foreground level.

\begin{figure}[t]
    \centering
    \includegraphics[width=0.95\linewidth]{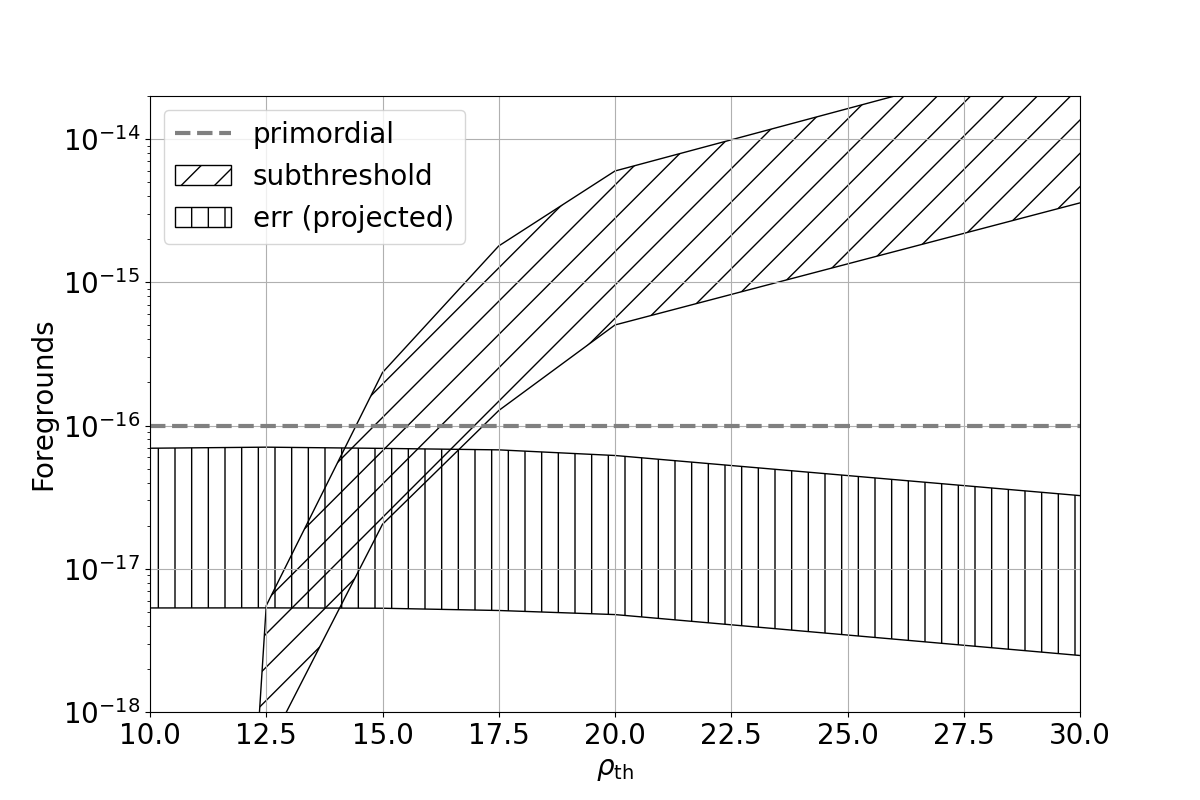}
    \caption{\ac{BNS} foreground components at the frequency of \qty{0.2}{Hz} with the uncertainty in the local merger rate density. The region hatched by vertical lines shows the subtraction errors with the projection procedure. The region hatched by tilted lines shows the subthreshold components. For both region, the upper and lower edges correspond to the upper and lower bound of the local merger rate density, respectively.}
    \label{fig: foregrounds with R0 uncertainty}
\end{figure}

The local merger rate density affects the foreground level rather than the mass spectrum models. The GWTC-4 population paper reported that the \ac{BNS} local merger rate density is estimated as \qtyrange{13}{170}{Gpc^{-3} yr^{-1}} with the \textsc{Simple Uniform} mass spectrum model.
Figure~\ref{fig: foregrounds with R0 uncertainty} shows the derived uncertainty in the subthreshold and projected error components as functions of \ac{SNR} threshold. The foreground level is approximately proportional to the local merger rate density. The widths of the hatched regions in Fig.~\ref{fig: foregrounds with R0 uncertainty} are about one order of magnitude, reflecting the similar order-of-magnitude uncertainty in the \ac{BNS} local merger rate density. The projected errors are still below the reference level of the primordial \ac{SGWB}. In contrast, the subthreshold component changes significantly. The uncertainty in the local merger rate density could affect the choice of the threshold \ac{SNR}.
The cutoff frequency of the unresolvable and the PE error components also depend on the local merger rate density. These frequencies vary from \qty{0.02}{Hz} to \qty{0.05}{Hz} and do not significantly affect the detectability of the primordial \ac{SGWB}.
For \acp{BBH}, the local merger rate density is uncertain only by a factor of \numrange{2}{3}. The results will not change significantly compared to the case of \acp{BNS}.

\begin{figure}[t]
    \centering
    \includegraphics[width=0.95\linewidth]{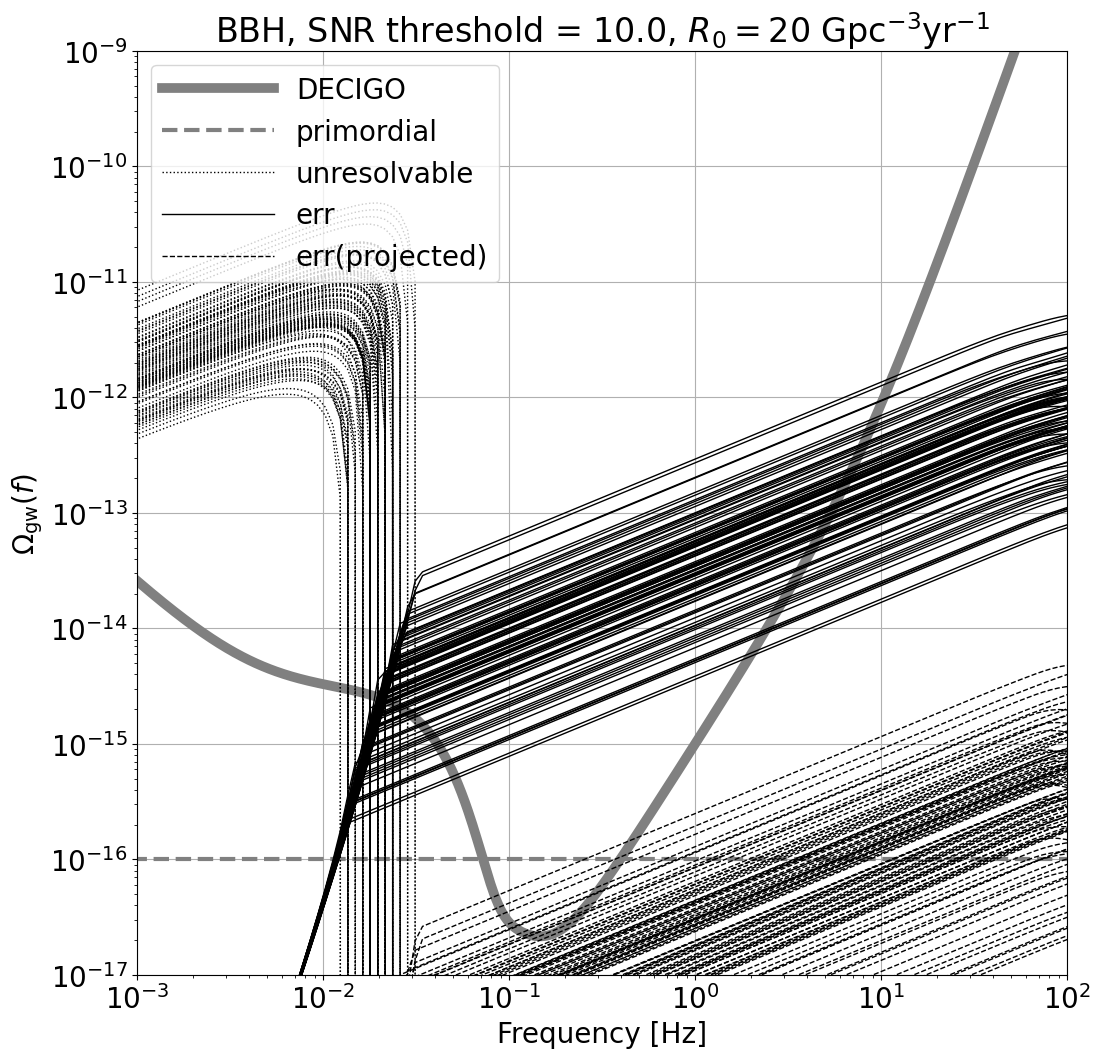}
    \caption{\ac{BBH} foreground components with the uncertainty in the merger rate history. We overplot \num{100} realizations of the foregrounds using the radomly sampled parameters characterizing the merger rate history. The solid and dashed lines are the PE errors without and with the projection, respectively. The dotted lines are the unresolvable components. The subthreshold components are not larger than \num{e-17}, which are out of the plot range.}
    \label{fig: foregrounds with merger rate history uncertainty}
\end{figure}

Lastly, we consider the uncertainty in the models of the merger rate history. GWTC4 stochastic paper~\cite{LIGOScientific:2025bgj} presents the uncertainty in the parameters $(\alpha, \beta, z_\mathrm{p})$ characterizing the merger rate density (see Eq.~\eqref{eq: merger rate density of bbh}). The 90 percent credible intervals of the uncertainties in the merger rate parameters are $\alpha \in [2.2, 4.3]$, $\beta \in [0.5, 9.6]$, and $z_\mathrm{p} \in [1.3, 3.8]$. Because the merger rate density is poorly constrained at redshift $z \gtrsim 2$, it leads to a large uncertinty in the foreground level. We randomly drow 100 samples from the uniform distributions on the above ranges and perform the same analysis to see how large the uncertainty in the foreground level is. \footnote{We choose the uniform distribution because the poseterior distribution or the posterior samples that are obtained by GWTC4 stochastic~\cite{LIGOScientific:2025bgj} are not publicly available.}
Figure~\ref{fig: foregrounds with merger rate history uncertainty} shows the \ac{BBH} foreground components with the uncertainty in the merger rate parameters. The subthreshold components have sharp cutoffs approximately between \qty{0.01}{Hz} and \qty{0.03}{Hz} that do not affect the detectability of the primordial \ac{SGWB}. The PE error components with the projection can vary by up to two orders of magnitude. In the case where the local merger rate density is high and the slope of the merger rate density at high redshift is gentle, the foregrounds can be comparable to the primordial \ac{SGWB}. For this case, the search method simultaneously searching the two \ac{SGWB} components could be useful~\cite{Martinovic:2020hru}.

For \acp{BNS}, the observational constraints are weak and not shown in~\cite{LIGOScientific:2025bgj}. Third generation ground-based detectors, such as Einstein Telescope~\cite{ET:2025xjr} and Cosmic Explorer~\cite{Evans:2021gyd}, will reach the \ac{BNS} mergers located around $z \gtrsim 2$. We might be able to obtain more detailed information on the merger rate density at high redshift.

There are scenarios where the merger rate history can not be modeled by the double power-law, Eq~\eqref{eq: merger rate density of bbh}. For example, \acp{BBH} originating the primordial fluctuation in the early Universe can largely enhance the number of \acp{BBH} in the distance redshift. The increase in the number of distant binaries leads the enhance the PE error component and the subthreshold component. These cases can be identified by alternative missions (e.g., B-DECIGO~\cite{Kawamura:2020pcg}), and we will be able to use the information to revisit the estimation of the foregrounds.

\section{Conclusion and discussion}
\label{sec: conclusion}

The ultimate configuration of DECIGO would be able to detect the primordial \ac{SGWB}.
Although it has high sensitivity, a large number of compact binaries forming the foreground hinders the detection of the primordial \ac{SGWB}.
Detecting the compact binaries as individual events and subtracting them is a crucial strategy for detecting the primordial \ac{SGWB} in DECIGO.
In this paper, we assessed the amplitude of the foreground by using the population model inferred from the latest \ac{LVK} observation.
We found that the projection scheme allows us to remove both the foregrounds of \ac{BBH} and \ac{BNS} and suppress them enough to detect the primordial \ac{SGWB}.

Regarding the feasibility of the subtraction, we compare the total number of source parameters and the number of data points available in the strain data.
Assuming the local merger rate of $\qty{100}{Gpc^{-3} yr^{-1}}$ and the population model of GWTC4, the total number of \acp{BNS} is about \num{8e5}.
Each of them is characterized by $\sim 10$ source parameters.
The total number of parameters to be estimated is about \num{8e6}.
On the other hand, assuming the observational period of \qty{3}{yr} and the sampling rate of \qty{1}{Hz}, we estimate the total number of data points in the strain data to be about \num{2e8}, which is larger than the total number of parameters to be estimated.
In this sense, it is feasible to estimate the best-fit parameters for all \acp{BNS}.
Practically, the computational cost for the parameter estimation is a serious problem.
Furthermore, in the deci Hz frequency band, the binary signals have long durations, which require a higher computational cost.
Even in \ac{LVK} analysis, the full-parameter estimation takes several hours for a \ac{BBH} merger signal and several days for a \ac{BNS} signal.
There are several sophisticated techniques for efficiently estimating the source parameters (heterodyned likelihood~\cite{Cornish:2010kf,Cornish:2021wxy}, relative binning~\cite{Zackay:2018qdy,Krishna:2023bug}, singular value decomposition of a template bank~\cite{Cannon:2010qh}, and simulation based inference~\cite{Cranmer:2019eaq,Green:2020hst,Delaunoy:2020zcu,Dax:2021myb,Dax:2023ozk,Raymond:2024xzj}).
Though, their application to the long-duration signals are limited yet, and the computational costs that are required for the foreground removal still be unexplored. It is important to estimate their computational cost and their ability for the foreground removal.


Several systematic uncertainties are not addressed in this paper. For example, if a binary has non-negligible eccentricity, the \ac{GW} signal is strongly modulated and has higher harmonics~\cite{Peters:1963ux}. The misaligned spin induces the orbital precession, which modulates the signal amplitude~\cite{Apostolatos:1994mx}. The environmental effects (e.g., gasses around bianries) can modify the waveforms. Besides these waveform systematics, the calibration error is another crucial factor. For the current ground-based detectors, the calibration errors are estimated to be a few percent in amplitude and a few degree in phase~\cite{Sun:2020wke,Sun:2021qcg,Aubin:2025dsv}. Neglecting these calibration errors induces the bias in the parameter estimation, and they would be more significant for signals with higher \acp{SNR}. Evaluating systematic errors due to the waveform mismodeling and the calibrations is essential for assessing the foreground.

\acknowledgements
The authors are greatful to Nobuyuki Kanda and Takahiro Tanaka for fruitful discussions, and to Kazunori Kohri and Xingjiang Zhu for their insightful comments.
This work is supported by JSPS KAKENHI Grant No.~JP23K13099, JP23H04502, and JP26K17138.

We used the software: \texttt{numpy}~\cite{Harris:2020xlr}, \texttt{scipy}~\cite{Virtanen:2019joe}, \texttt{matplotlib}~\cite{Hunter:2007ouj}, \texttt{astropy}~\cite{Astropy:2013muo,Astropy:2018wqo,Astropy:2022ucr}.

\appendix
\section{Cutler and Harms' projection strategy}
\label{appendix}

In this appendix, we explain how we assessed the subtraction errors in Eqs.~\eqref{eq: omega_err} and~\eqref{eq: omega_err_p}.

\subsection{Best-fit waveform}
\label{appendix a1: best-fit}

The log-likelihood that the strain data $d$ contains the \ac{GW} signal $h(\theta)$ is given by
\begin{equation}
    \Lambda(d \mid \theta) = - \frac{1}{2} \langle d - h(\theta) \mid d - h(\theta) \rangle\,.
    \label{eq: def log likelihood}
\end{equation}
The bracket notation, $\langle a \mid b \rangle$, is the noise-weighted inner product defined by
\begin{equation}
    \langle a \mid b \rangle = 4\Re \int^\infty_0 \dd{f} \frac{\tilde{a}^\ast(f) \tilde{b}(f)}{S_\mathrm{n}(f)}\,.
\end{equation}
We assume that \ac{GW} signal $\hat{h} \coloneqq h(\hat{\theta})$ with the true parameter exists in the strain data $d$, written by
\begin{equation}
    d = \hat{h} + n\,,
\end{equation}
with the detector noise $n$.
The best-fit parameter, $\theta_\bestfit$, is defined as the parameter $\theta$ where the log-likelihood becomes its maximum.
From the definition of the log-likelihood~\eqref{eq: def log likelihood}, we get
\begin{equation}
    \langle \partial_i h_\bestfit \mid \hat{h} - h_\bestfit \rangle + \langle \partial_i h_\bestfit \mid n \rangle = 0\,,
    \label{eq: def bestfit}
\end{equation}
where $\partial_i$ is the partial derivative with respect to $i$ th parameter of $\theta$.
The bestfit waveform is denoted by $h_\bestfit \coloneqq h(\theta_\bestfit)$.
We denote the deviation of the best-fit parameter from the true value by $\Delta\theta$,
\begin{equation}
    \Delta\theta = \theta_\bestfit - \hat{\theta}\,.
\end{equation}
We define the residual $\delta h$ that will be left after subtracting the bestfit waveform $h_\bestfit$ from the true signal $\hat{h}$, that is
\begin{equation}
    \delta h = \hat{h} - h_\bestfit\,.
\end{equation}
Assuming the deviation of the best-fit parameter from the true value is small enough, we perform the Taylor expansion of $\delta h$ around $\Delta\theta=0$ as
\begin{align}
    \delta h &= \hat{h} - \Bqty{ \hat{h} + \Delta\theta_i \partial_i h_\bestfit + \frac{1}{2} \Delta\theta_i \Delta\theta_j \partial_i \partial_j h_\bestfit + \cdots} \notag\\
    &= - \Delta\theta_i \partial_i h_\bestfit - \frac{1}{2} \Delta\theta_i \Delta\theta_j \partial_i \partial_j h_\bestfit + \cdots\,.
    \label{eq: delta h}
\end{align}
Here, we use the Einstein summation convention.
Substituting Eq.~\eqref{eq: delta h} into Eq.~\eqref{eq: def bestfit}, we get
\begin{equation}
    0 = - \langle \partial_i h_\bestfit \mid \partial_j h_\bestfit \rangle \Delta\theta_j + \langle \partial_i h_\bestfit \mid n \rangle + O(\Delta\theta^2)\,.
\end{equation}
The deviation $\Delta\theta_i$ can be written by
\begin{equation}
    \Delta\theta_i = (\Gamma_\bestfit^{-1})_{ij} \langle \partial_j h_\bestfit \mid n\rangle\,,
\end{equation}
where $\Gamma$ is the Fisher matrix,
\begin{equation}
    \Gamma_{ij}(\theta) = \langle \partial_i h(\theta) \mid \partial_j h(\theta) \rangle\,,
\end{equation}
and $(\Gamma_\bestfit)_{ij} = \Gamma_{ij}(\theta_\bestfit)$.
The covariance of $\Delta\theta$ is obtained by
\begin{equation}
    \overline{\Delta\theta_i \Delta\theta_j} = (\Gamma^{-1}_\bestfit)_{ij}\,.
\end{equation}
We get a rough estimation of the power of the residual $\delta h$,
\begin{equation}
    \frac{\overline{ \langle \delta h \mid \delta h \rangle}}{\langle h \mid h \rangle}
    = \frac{\overline{\Delta\theta_i \Delta\theta_j} \langle \partial_i h_\bestfit \mid \partial_j h_\bestfit \rangle}{\langle h \mid h \rangle}
    = \frac{N_\mathrm{p}}{\rho^2}\,.
    \label{eq: dh^2/h^2}
\end{equation}
up to the fourth order of $\Delta\theta$.
Here, $N_\mathrm{p}$ is the number of source parameters.
We use Eq.~\eqref{eq: dh^2/h^2} for assessing the residual components $\Omega_\mathrm{err}$ in Eq.~\eqref{eq: omega_err}.

\subsection{Projection}
\label{appendix a2: projection}

We define the projection operator $P$ and the projected residual $\delta^\mathrm{P} h$ by
\begin{equation}
    \delta^\mathrm{P} h = P[\delta h] = \delta h - (\partial_i h_\bestfit) (\Gamma^{-1}_\bestfit)_{ij} \langle \partial_j h_\bestfit \mid \delta h\rangle\,.
    \label{eq: projection op}
\end{equation}
Substituting the Taylor expansion of $\delta h$ with respect to the $\Delta\theta$ and keeping up to the second order of $\Delta\theta$, we get
\begin{align}
    \delta^\mathrm{P} h
    &= - \frac{1}{2} \Delta\theta_i \Delta\theta_j \partial_i \partial_j h_\bestfit \\\notag
    &\quad + \Delta\theta_i \Delta\theta_j (\partial_k h_\bestfit) (\Gamma^{-1}_\bestfit)_{k\ell} \langle \partial_\ell h_\bestfit \mid \partial_i \partial_j h_\bestfit \rangle\,.
\end{align}
We get the power of the projected residual as
\begin{equation}
    \langle \delta^\mathrm{P} h \mid \delta^\mathrm{P} h\rangle
    = \frac{1}{4} \Delta\theta_i \Delta\theta_j \Delta\theta_k \Delta\theta_\ell
    \langle \partial_i \partial_j h_\bestfit \mid \partial_k \partial_\ell h_\bestfit \rangle
\end{equation}
Evaluating this with the rough estimation where $\partial^2 h \sim (\partial h)^2 / h \sim \Gamma / h$ and $(\Delta\theta)^2 \sim \Gamma^{-1}$, we get
\begin{equation}
    \frac{\overline{ \langle \delta^\mathrm{P} h \mid \delta^\mathrm{P} h \rangle }}{\langle h \mid h \rangle}
    \sim \frac{N_\mathrm{p}^2}{\rho^4}\,.
    \label{eq: dhp^2/h^2}
\end{equation}
We use this in Eq.~\eqref{eq: omega_err_p}.

\bibliography{reference}
\end{document}